\documentclass[aps,pre,
onecolumn,
groupedaddress,showpacs]{revtex4-2}

\usepackage{threeparttable}
\usepackage{epsfig}
\usepackage{epstopdf}

\usepackage{graphicx}

\usepackage{amsmath}
\usepackage[T1]{fontenc}
\usepackage{caption}
\usepackage{float} 
\usepackage{dcolumn}
\usepackage{hyperref}
\usepackage{amsthm}
\usepackage{amssymb}

\usepackage{stackengine}
\newcommand\xrowht[2][0]{\addstackgap[.5\dimexpr#2\relax]{\vphantom{#1}}}
\usepackage{booktabs}
\usepackage{xparse}
\usepackage{tikz}
\usetikzlibrary{calc}

\tikzset{Arrow Style/.style={text=black, font=\boldmath}}

\newcommand*{\XShift}{0.5em}
\newcommand*{\YShift}{0.5ex}

\NewDocumentCommand{\DrawArrow}{s O{} m m m}{%
    \begin{tikzpicture}[overlay,remember picture]
        \draw[->, thick, Arrow Style, #2] 
                ($(#3.west)+(\XShift,\YShift)$) -- 
                ($(#4.east)+(-\XShift,\YShift)$)
        node [midway, above] {#5};
    \end{tikzpicture}%
}

\def\e{\mbox{e}}
\def\d{\mbox{d}}
\def\i{\small{\mbox{i}}}

\begin{document}


\title{Summation formulas generated by Hilbert space eigenproblem}
\author{Petar Mali$^{1}$} \email[]{petar.mali@df.uns.ac.rs} \author{Sonja Gombar$^{1}$} \author{Slobodan Rado\v sevi\' c$^{1,2}$}  \author{Milica Rutonjski$^{1}$} \author{Milan Panti\' c$^{1}$} \author{Milica Pavkov-Hrvojevi\' c$^{1}$}
\affiliation{$^1$ Department of Physics, Faculty of Sciences, University of Novi Sad,
Trg Dositeja Obradovi\' ca 4, 21000 Novi Sad, Serbia\\
$^2$ Universidad de Ja\'{e}n, Departamento de Matem\'{a}ticas, Spain}

\date{\today}


\begin{abstract}
We demonstrate that certain classes of Schl\" omilch-like infinite series and series that include generalized hypergeometric functions can be calculated in closed form starting from a simple quantum model of a particle trapped inside an infinite potential well and using principles of quantum mechanics. We provide a general framework based on the Hilbert space eigenproblem that can be applied to different exactly solvable quantum models. Obtaining series from normalization conditions in well-defined quantum problems secures their convergence.  
\end{abstract}
\maketitle

\section{Introduction}
\label{intro}
George Polya noted that: "Mathematical problems are often inspired by nature, or rather by our interpretation of nature, the physical sciences. Also, the solution of a mathematical problem may be inspired by nature; physics provides us with clues with which, left alone, we had very little chance to provide ourselves." \cite{polya} In other words, neither is physics merely an area of application of mathematical expressions nor is mathematics just an observant to physics, providing her with a necessary set of tools - they are two nontrivial disciplines standing together as important sidekicks to one another, intertwining in such delicate and eligible manner to present key pieces in understanding the ultimate foundations of either of them. 
This paper aims to provide yet another illustration of what Polya named physical mathematics, the discipline in which mathematical results are obtained starting from nontrivial physical problems. Some examples of this kind of reasoning, along with those identified by Polya in \cite{polya}, include \cite{levi,uspeh,nakahara,random,mali}. 

In the present paper, a simple quantum mechanical model is exploited to obtain new infinite sums including products of Bessel functions (Schl\" omilch-like infinite series \cite{miller}), as well as those containing generalized hypergeometric functions, which, up to our knowledge, have not been calculated yet. One of the simplest exactly solvable quantum mechanical models, used also in \cite{mali}, describes a particle of mass $m$ moving in the infinite potential well. The Hamiltonian of that problem can be specified as $\hat{H}=-\frac{\hbar^2}{2m}\frac{\mathrm{d}^2}{\mathrm{d} x^2}+V(x)$,
where
\begin{equation}
 V(x) =
  \begin{cases}
                                   0, & x \in (0,a) \\
                                   \infty, & x \notin (0,a)
 \end{cases}
\end{equation}
and $a$ is the well width. The normalized eigenstates of Hamiltonian are  $\psi_n(x)=\sqrt{\frac{2}{a}}\sin\frac{n \pi x}{a}, n=1,2,3...$ and the system can be prepared in an arbitrary state described by the normalized two-time differentiable function $\psi(x)$ which is an element of $L^2([0,a])$ and satisfies $\psi(0)=\psi(a)=0$. Since eigenstates of Hamiltonian form an orthonormal basis, any state $\psi(x)$ can be written as an infinite linear combination of eigenstates of Hamiltonian
\begin{equation}
    \psi(x)=\sum^{\infty}_{n=1}C_n \psi_n(x), \label{linearnakom}
\end{equation}
where 
\begin{equation}
C_n=\int^{a}_0 \psi(x)\psi_n(x)\mathrm{d}x.
\end{equation}
It is clear that for certain states $\psi(x)$ (\ref{linearnakom}) reduces to a finite series. These states, however, will not be considered in this paper. The states of the polynomial type $\psi(x)=Cx^{j}(a-x)^{l}Q_N(x)$, where $j$, $l$ are integers, $Q_N$ is arbitrary polynomial of degree $N$, and $C$ is the normalization constant, were used in \cite{mali} to evaluate analytically different Riemann zeta $\zeta(p)$ and related functions for positive even arguments $p \geq 2$. In the present paper, we will use non-polynomial and non-trigonometric states in order to obtain new infinite sums. 

The paper is organized as follows: the results are introduced in Sec. II, and Sec. III concludes the paper.  
\section{Results}
\label{Results}
The simplest non-polynomial and non-trigonometric wave function that satisfies the boundary conditions in the infinite well is given by $\psi(x) = Cx^{\alpha}(a-x)^{\beta}$, where $C$ represents the normalization constant, and $\alpha$ or $\beta$ are positive non-integer rational numbers. The wave function is normalized by $\int^a_0|\psi(x)|^2\d x=1$. Let us emphasize that only the functions of this type, containing zero nodes, will be the subject of interest in the present paper.
\subsection{Non-polynomial and non-trigonometric wave functions for $\alpha=\beta$} \label{A}
 Let us specify that $\alpha$ and $\beta$ are both equal to $1/2$. In that case, coefficients in the infinite series (\ref{linearnakom}) of the function $\psi(x)=\sqrt{\frac{6}{a^3}}x^{\frac{1}{2}}(a-x)^{\frac{1}{2}}$ are
\begin{equation}
C_n=\int^{a}_0 \psi(x)\psi_n(x)\mathrm{d}x=\frac{\sqrt{3}}{n}J_1\left(\frac{n\pi}{2}\right)\sin\left(\frac{n\pi}{2}\right), \label{prvkon}
\end{equation}
where $J_1(x)$ is the Bessel function of the first kind. From the condition $\sum^{\infty}_{n=1}|C_n|^2=1$, we obtain
\begin{equation}
\sum^{\infty}_{n=1}\frac{J_1^2\left(\frac{n\pi}{2}\right)\sin^2\left(\frac{n\pi}{2}\right)}{n^2}=\frac{1}{3}. \label{prvsum}
\end{equation}
Since only the odd terms give a contribution to the infinite series, it can be written as
\begin{equation}  
\sum^{\infty}_{k=0}\frac{J_1^2\left(\frac{(2k+1)\pi}{2}\right)}{(2k+1)^2}=\frac{1}{3}.
\end{equation}

In (\ref{prvkon}) we used \cite{GRRI,erdel} 
\begin{equation}
\int^u_0 x^{\mu-1}(u-x)^{\mu-1}\sin(\tilde{a}x)\d x=\sqrt{\pi}\left(\frac{u}{\tilde{a}}\right)^{\mu-\frac{1}{2}}\sin \left(\frac{\tilde{a}u}{2}\right)\Gamma(\mu)J_{\mu-\frac{1}{2}}\left(\frac{\tilde{a}u}{2}\right), \label{kljucno}
\end{equation}
where $\mathrm{Re}\,\mu>0$. For more details see \ref{app}.
Bearing in mind that the gamma function can be calculated analytically only for integer and half-integer arguments, in this section we will focus only on half-integer values of $\alpha$ and $\beta$. One should note that for integer values of the argument of the gamma function the wave function will be of polynomial type which is investigated in \cite{mali}.
Now we proceed to the next logical choice of parameters $\alpha=\beta=\frac{3}{2}$, i.e.
$\psi(x)=2\sqrt{\frac{35}{a^7}}x^{\frac{3}{2}}(a-x)^{\frac{3}{2}}.$
From that we obtain infinite sum
\begin{equation}
\sum^{\infty}_{n=1}\frac{J_2^2\left(\frac{n\pi}{2}\right)\sin^2\left(\frac{n\pi}{2}\right)}{n^4}=\frac{2\pi^2}{315},
\end{equation}
which is equivalent to
\begin{equation}
\sum^{\infty}_{k=0}\frac{J_2^2\left(\frac{(2k+1)\pi}{2}\right)}{(2k+1)^4}=\frac{2\pi^2}{315}.
\end{equation}
Furthermore, for $\alpha=\beta=\frac{5}{2}$ we start the procedure with the wave function
$\psi(x)=6\sqrt{\frac{77}{a^{11}}}x^{\frac{5}{2}}(a-x)^{\frac{5}{2}}$
to obtain
\begin{equation}
\sum^{\infty}_{n=1}\frac{J_3^2\left(\frac{n\pi}{2}\right)\sin^2\left(\frac{n\pi}{2}\right)}{n^6}=\frac{8\pi^4}{155925},
\end{equation}
which is equivalent to
\begin{equation}
\sum^{\infty}_{k=0}\frac{J_3^2\left(\frac{(2k+1)\pi}{2}\right)}{(2k+1)^6}=\frac{8\pi^4}{155925}.
\end{equation}
Following this type of procedure can lead us to infinitely many infinite sums, some of which can be found in Table \ref{tabelasume}. Note that for any pair $(\alpha,\beta)$, $\alpha=\beta$, one infinite sum is obtained, which is completely independent of the other infinite sums. Therefore, in this case, the order in which we compute a particular infinite sum is not important. 

\begin{widetext}
\begin{center}
\begin{table}
 
\caption{List of the infinite series $\sum^{\infty}_{k=0}\frac{J_p^2\left(\frac{(2k+1)\pi}{2}\right)}{(2k+1)^{2p}}$  that can be calculated for $\alpha=\beta$.}
\begin{threeparttable}
 \begin{tabular}{|c |c| c |c|} 

 \hline\hline
 $\alpha=\beta$ & $\psi(x)$  & p & sum of the infinite series \\ [0.5ex] 
 \hline\hline
 $\frac{1}{2}$ & $\sqrt{\frac{6}{a^3}}x^{\frac{1}{2}}(a-x)^{\frac{1}{2}}$  & $1$ &  $\frac{1}{3}$ \\ \hline
 
 $\frac{3}{2}$ & $2\sqrt{\frac{35}{a^7}}x^{\frac{3}{2}}(a-x)^{\frac{3}{2}}$ & $2$  &  $\frac{2\pi^2}{315}$   \\ \hline

 $\frac{5}{2}$ & $6\sqrt{\frac{77}{a^{11}}}x^{\frac{5}{2}}(a-x)^{\frac{5}{2}}$ & $3$  &  $\frac{8\pi^4}{155925}$   \\ \hline

 $\frac{7}{2}$ & $6\sqrt{\frac{1430}{a^{15}}}x^{\frac{7}{2}}(a-x)^{\frac{7}{2}}$ & $4$  &  $\frac{16\pi^6}{70945875}$   \\ \hline

 $\frac{9}{2}$ & $2\sqrt{\frac{230945}{a^{19}}}x^{\frac{9}{2}}(a-x)^{\frac{9}{2}}$ & $5$  &  $\frac{128\pi^8}{206239658625}$   \\ \hline

 $\frac{11}{2}$ & $2\sqrt{\frac{4056234}{a^{23}}}x^{\frac{11}{2}}(a-x)^{\frac{11}{2}}$ & $6$  &  $\frac{256\pi^{10}}{219150261254925}$   \\ \hline

 $\frac{13}{2}$ & $30\sqrt{\frac{312018}{a^{27}}}x^{\frac{13}{2}}(a-x)^{\frac{13}{2}}$ & $7$  &  $\frac{1024\pi^{12}}{641014514170655625}$   \\ \hline

 $\frac{15}{2}$ & $12\sqrt{\frac{33393355}{a^{31}}}x^{\frac{15}{2}}(a-x)^{\frac{15}{2}}$ & $8$  &  $\frac{2048\pi^{14}}{1234868674798755871875}$   \\ \hline

  $\frac{17}{2}$ & $30\sqrt{\frac{90751353}{a^{35}}}x^{\frac{17}{2}}(a-x)^{\frac{17}{2}}$ & $9$  &  $\frac{32768\pi^{16}}{24246646429673571544265625}$   \\ \hline

  $\frac{19}{2}$ & $30\sqrt{\frac{1531628098}{a^{39}}}x^{\frac{19}{2}}(a-x)^{\frac{19}{2}}$ & $10$  &  $\frac{65536\pi^{18}}{73863367240262256781014515625}$   \\ \hline

 \hline\hline 
\end{tabular}

\end{threeparttable}
\label{tabelasume}
\end{table}
\end{center}
\end{widetext}

\subsection{Non-polynomial and non-trigonometric wave functions for $\alpha \neq \beta$}
For $\alpha \neq \beta$ we employ \cite{GRRI,erdel} 
\begin{widetext}
\begin{align}
\int^u_0 x^{\nu-1}(u-x)^{\mu-1}\sin(\tilde{a}x)\d x&=\frac{u^{\mu+\nu-1}}{2\mathrm{i}}B(\mu,\nu)[_1\hspace{-0.5mm}F_1(\nu;\mu+\nu;\mathrm{i}\tilde{a}u)-_1\hspace{-1.3mm}F_1(\nu;\mu+\nu;-\mathrm{i}\tilde{a}u)],\,\, \nonumber \\ a&>0, \mathrm{Re}\,\mu>0, \mathrm{Re}\,\nu>-1, \nu \neq 0.
\label{kljucno2}
\end{align}
\end{widetext}
For more details see \ref{app}. 
Let's start with the case $\alpha=\frac{3}{2}$, $\beta=\frac{1}{2}$, i.e. $\psi(x)=2\sqrt{\frac{5}{a^5}}x^{\frac{3}{2}}(a-x)^{\frac{1}{2}}.$ Therefrom, employing (\ref{kljucno2}) leads to 
\begin{equation}
    C_n=\frac{\sqrt{10}}{\i}B\left(\frac{3}{2},\frac{5}{2}\right)\left[_1\hspace{-0.05cm}F_1\left(\frac{5}{2};4;\i n\pi\right)-_{1}\hspace{-0.12cm}F_1\left(\frac{5}{2};4;-\i n\pi\right)\right],
    \end{equation}
    where 
    \begin{equation}
   _1\hspace{-0.05cm}F_1\left(\frac{5}{2};4;\pm \i n\pi\right)=\frac{4\e^{\pm \frac{\mathrm{i} n\pi}{2}}(\mp\i n\pi J_0(\frac{n\pi}{2}) \pm 4\i J_1(\frac{n\pi}{2})+n\pi J_1(\frac{n\pi}{2}))}{n^2\pi^2}.     
    \end{equation}
    In expression for 
    $|C_n|^2$ we use the recurrence relation for the Bessel functions
    $J_0(\frac{n\pi}{2})=\frac{4}{n\pi}J_1(\frac{n\pi}{2})-J_2(\frac{n\pi}{2}),$ and after that, from $\sum^{\infty}_{n=1}|C_n|^2=1$, we get
    \begin{equation}
\sum^{\infty}_{n=1}\frac{J_2^2(\frac{n\pi}{2})\cos^2(\frac{n\pi}{2})}{n^2}+\sum^{\infty}_{n=1}\frac{J_1^2(\frac{n\pi}{2})\sin^2(\frac{n\pi}{2})}{n^2}=\frac{2}{5}.  
\end{equation}
Having already calculated the second sum (see (\ref{prvsum})),  we obtain 
\begin{equation}
  \sum^{\infty}_{n=1}\frac{J_2^2(\frac{n\pi}{2})\cos^2(\frac{n\pi}{2})}{n^2}=\frac{1}{15},  
\end{equation}
i.e.
\begin{equation}
\sum^{\infty}_{k=1}\frac{J_2^2(k \pi)}{(2k)^2}=\frac{1}{15},  
\end{equation}
Following the same procedure for $\alpha=\frac{1}{2}$, $\beta=\frac{5}{2}$, i.e. $\psi(x)=\sqrt{\frac{42}{a^7}}x^{\frac{1}{2}}(a-x)^{\frac{5}{2}}$,
we obtain
\begin{equation}
\sum^{\infty}_{n=1}\frac{J_1(\frac{n\pi}{2})J_2(\frac{n\pi}{2})\sin^2 \frac{n\pi}{2}}{n^3}=\frac{2\pi}{45},
\end{equation}
which is equivalent to
\begin{equation}
\sum^{\infty}_{k=0}\frac{J_1\left(\frac{(2k+1)\pi}{2}\right)J_2\left(\frac{(2k+1)\pi}{2}\right)}{(2k+1)^3}=\frac{2\pi}{45}.
\end{equation}
\textit{Note that, unlike the case of $\alpha=\beta$, for $\alpha\neq\beta$ we must proceed with the calculations in a pyramidal manner, from lower to higher values of $\alpha$ and $\beta$, since for higher values of $\alpha$ and $\beta$ the expressions contain infinite series obtained from the lower ones.}
Moreover, for $\alpha=\frac{3}{2}$, $\beta=\frac{5}{2}$, i.e. $\psi(x)=6\sqrt{\frac{14}{a^9}}x^{\frac{3}{2}}(a-x)^{\frac{5}{2}}$,  with recurrence relation
$J_1(\frac{n \pi}{2})=\frac{8}{n \pi}J_2(\frac{n \pi}{2})-J_3(\frac{n \pi}{2})$ we get
\begin{equation}
    \sum^{\infty}_{n=1}\frac{J_3^2(\frac{n \pi}{2})\cos^2 (\frac{n\pi}{2})}{n^4}=\frac{2\pi^2}{2835},
\end{equation}
which is equivalent to
\begin{equation}
    \sum^{\infty}_{k=1}\frac{J_3^2(k \pi)}{(2k)^4}=\frac{2\pi^2}{2835}.
\end{equation}
In the case of state $\psi(x)=6\sqrt{\frac{2}{a^9}}x^{\frac{1}{2}}(a-x)^{\frac{7}{2}}$, after applying $J_0(\frac{n\pi}{2})=\frac{4}{n\pi}J_1(\frac{n\pi}{2})-J_2(\frac{n\pi}{2})$ we get the equation with multiple unknown infinite sums
\begin{eqnarray}
    \frac{4}{9}&=&\frac{9}{\pi^2}\textcolor[rgb]{1,0,0}{\sum^{\infty}_{n=1}\frac{J_1^2(\frac{n \pi}{2})\cos^2 (\frac{n \pi}{2})}{n^4}}+4 \sum^{\infty}_{n=1}\frac{J_1^2(\frac{n \pi}{2})\sin^2 (\frac{n \pi}{2})}{n^2}+\frac{576}{\pi^4}\textcolor[rgb]{1,0,0}{ \sum^{\infty}_{n=1}\frac{J_2^2(\frac{n \pi}{2})\cos^2 (\frac{n \pi}{2})}{n^6}}-\frac{96}{\pi^2} \textcolor[rgb]{1,0,0}{\sum^{\infty}_{n=1}\frac{J_2^2(\frac{n \pi}{2})\cos^2 (\frac{n \pi}{2})}{n^4}} \nonumber\\
&+&4\sum^{\infty}_{n=1}\frac{J_2^2(\frac{n \pi}{2})\cos^2 (\frac{n \pi}{2})}{n^2}+\frac{81}{\pi^2}\sum^{\infty}_{n=1}\frac{J_2^2(\frac{n \pi}{2})\sin^2 (\frac{n \pi}{2})}{n^4}-\frac{144}{\pi^3}\textcolor[rgb]{1,0,0}{\sum^{\infty}_{n=1}\frac{J_1(\frac{n \pi}{2})J_2(\frac{n \pi}{2})\cos^2 (\frac{n \pi}{2})}{n^5}}\nonumber \\
&+&\frac{12}{\pi}\textcolor[rgb]{1,0,0}{\sum^{\infty}_{n=1}\frac{J_1(\frac{n \pi}{2})J_2(\frac{n \pi}{2})\cos^2 (\frac{n \pi}{2})}{n^3}}
-\frac{36}{\pi}\sum^{\infty}_{n=1}\frac{J_1(\frac{n \pi}{2})J_2(\frac{n \pi}{2})\sin^2 (\frac{n \pi}{2})}{n^3}.
\label{jedanzasvi}\end{eqnarray}
With red color we mark the infinite sums that have not been calculated yet. One possibility to proceed is to write different wave functions and try to eliminate marked infinite sums by solving the system of equations. However, in this case it does not simplify the problem since more unknown infinite sums appear (even for $\psi(x)=6\sqrt{\frac{2}{a^9}}x^{\frac{7}{2}}(a-x)^{\frac{1}{2}}$ we also get (\ref{jedanzasvi})) and due to that we were unable to calculate new infinite sums. The second possibility is the further use of recurrence relations for the Bessel functions. After trying both methods, we conclude that for $|\alpha-\beta|>2$ the application of recurrence relations for the Bessel functions does not allow obtaining new sums analytically. Therefore, we carry on with the research for $|\alpha-\beta|\leq 2$. 

For some combinations of $\alpha$ and $\beta$ that satisfy mentioned condition we need to use the recurrence relation $J_{k+1}(x)+J_{k-1}(x)=\frac{2k}{x}J_k(x)$ multiple times in order to avoid equations with many unknown infinite sums. For instance, in case of $\alpha=\frac{5}{2}$, $\beta=\frac{9}{2}$,
$\psi(x)=\sqrt{\frac{30030}{a^{15}}}x^{\frac{5}{2}}(a-x)^{\frac{9}{2}}$
we used $J_0(\frac{n \pi}{2})=\frac{4}{n \pi}J_1(\frac{n \pi}{2})-J_2(\frac{n \pi}{2})$, $J_1(\frac{n \pi}{2})=\frac{8}{n \pi}J_2(\frac{n \pi}{2})-J_3(\frac{n \pi}{2})$, and $J_2(\frac{n \pi}{2})=\frac{12}{n \pi}J_3(\frac{n \pi}{2})-J_4(\frac{n \pi}{2})$
to obtain infinite sum
\begin{equation}
 \sum^{\infty}_{n=1}\frac{J_3(\frac{n \pi}{2})J_4(\frac{n \pi}{2})\sin^2(\frac{n \pi}{2})}{n^7}=\frac{16\pi^5}{4729725}, 
\end{equation}
which we can also write as
\begin{equation}
 \sum^{\infty}_{k=0}\frac{J_3(\frac{(2k+1) \pi}{2})J_4(\frac{(2k+1) \pi}{2})}{(2k+1)^7}=\frac{16\pi^5}{4729725}. 
\end{equation}
The sums obtained for $|\alpha-\beta|=2$ and for $|\alpha-\beta|=1$ are shown in Table \ref{tabelasume2}
and Table \ref{tabelasume3}, respectively. 
\begin{center}
\begin{table}
 
\caption{List of the infinite series $\sum^{\infty}_{k=0}\frac{J_p\left(\frac{(2k+1)\pi}{2}\right)J_q\left(\frac{(2k+1)\pi}{2}\right)}{(2k+1)^{p+q}}$  that can be calculated for $|\alpha-\beta|=2$.}
\begin{threeparttable}
 \begin{tabular}{|c |c| c |c|} 

 \hline\hline
 $\alpha;\beta$ & $\psi(x)$  & p;q &  sum of the infinite series \\ [0.5ex] 
 \hline\hline
 $\frac{1}{2}; \frac{5}{2}$ & $\sqrt{\frac{42}{a^7}}x^{\frac{1}{2}}(a-x)^{\frac{5}{2}}$  & $1;2$ &  $\frac{2\pi}{45}$ \\ \hline

 $\frac{3}{2}; \frac{7}{2}$ & $2\sqrt{\frac{330}{a^{11}}}x^{\frac{3}{2}}(a-x)^{\frac{7}{2}}$  & $2;3$ &  $\frac{8\pi^3}{14175}$ \\ \hline
 
 $\frac{5}{2};\frac{9}{2}$ & $\sqrt{\frac{30030}{a^{15}}}x^{\frac{5}{2}}(a-x)^{\frac{9}{2}}$ & $3;4$  &  $\frac{16\pi^5}{4729725}$   \\ \hline

 $\frac{7}{2};\frac{11}{2}$ & $12\sqrt{\frac{4199}{a^{19}}}x^{\frac{7}{2}}(a-x)^{\frac{11}{2}}$ & $4;5$  &  $\frac{128\pi^7}{10854718875}$   \\ \hline

 $\frac{9}{2};\frac{13}{2}$ & $2\sqrt{\frac{2860165}{a^{23}}}x^{\frac{9}{2}}(a-x)^{\frac{13}{2}}$ & $5;6$  &  $\frac{256\pi^9}{9528272228475}$   \\ \hline

 $\frac{11}{2};\frac{15}{2}$ & $12\sqrt{\frac{1448655}{a^{27}}}x^{\frac{11}{2}}(a-x)^{\frac{15}{2}}$ & $6;7$  &  $\frac{1024\pi^{11}}{23741278302616875}$   \\ \hline

 $\frac{13}{2};\frac{17}{2}$ & $15\sqrt{\frac{16500246}{a^{31}}}x^{\frac{13}{2}}(a-x)^{\frac{17}{2}}$ & $7;8$  &  $\frac{2048\pi^{13}}{39834473380605028125}$   \\ \hline

 $\frac{15}{2};\frac{19}{2}$ & $20\sqrt{\frac{162397158}{a^{35}}}x^{\frac{15}{2}}(a-x)^{\frac{19}{2}}$ & $8;9$  &  $\frac{32768\pi^{15}}{692761326562102044121875}$   \\ \hline

  $\frac{17}{2};\frac{21}{2}$ & $6\sqrt{\frac{31179571995}{a^{39}}}x^{\frac{17}{2}}(a-x)^{\frac{21}{2}}$ & $9;10$  &  $\frac{65536\pi^{17}}{1893932493340057866179859375}$   \\ \hline

 \hline\hline 
\end{tabular}

\end{threeparttable}
\label{tabelasume2}
\end{table}
\end{center}

\begin{center}
\begin{table}
 
\caption{List of the infinite series $\sum^{\infty}_{k=1}\frac{J_p^2 (k \pi)}{(2k)^{2p-2}}$  that can be calculated for $|\alpha-\beta|=1$.}
\begin{threeparttable}
 \begin{tabular}{|c |c| c |c|} 

 \hline\hline
 $\alpha;\beta$ & $\psi(x)$  & p & sum of the infinite series \\ [0.5ex] 
 \hline\hline
 $\frac{1}{2}; \frac{3}{2}$ & $\sqrt{\frac{20}{a^{5}}}x^{\frac{1}{2}}(a-x)^{\frac{3}{2}}$  & $2$ &  $\frac{1}{15}$ \\ \hline

 $\frac{3}{2}; \frac{5}{2}$ & $6\sqrt{\frac{14}{a^{9}}}x^{\frac{3}{2}}(a-x)^{\frac{5}{2}}$  & $3$ &  $\frac{2\pi^2}{2835}$ \\ \hline
 
 $\frac{5}{2};\frac{7}{2}$ & $6\sqrt{\frac{286}{a^{13}}}x^{\frac{5}{2}}(a-x)^{\frac{7}{2}}$ & $4$  &  $\frac{8\pi^4}{2027025}$   \\ \hline

 $\frac{7}{2};\frac{9}{2}$ & $4\sqrt{\frac{12155}{a^{17}}}x^{\frac{7}{2}}(a-x)^{\frac{9}{2}}$ & $5$  &  $\frac{16\pi^6}{1206079875}$   \\ \hline

 $\frac{9}{2};\frac{11}{2}$ & $2\sqrt{\frac{881790}{a^{21}}}x^{\frac{9}{2}}(a-x)^{\frac{11}{2}}$ & $6$  &  $\frac{128 \pi^8}{4331032831125}$   \\ \hline

 $\frac{11}{2};\frac{13}{2}$ & $20\sqrt{\frac{156009}{a^{25}}}x^{\frac{11}{2}}(a-x)^{\frac{13}{2}}$ & $7$  &  $\frac{256 \pi^{10}}{5478756531373125}$   \\ \hline

 $\frac{13}{2};\frac{15}{2}$ & $12\sqrt{\frac{7540435}{a^{29}}}x^{\frac{13}{2}}(a-x)^{\frac{15}{2}}$ & $8$  &  $\frac{1024 \pi^{12}}{18589420910949013125}$   \\ \hline

 $\frac{15}{2};\frac{17}{2}$ & $12\sqrt{\frac{129644790}{a^{33}}}x^{\frac{15}{2}}(a-x)^{\frac{17}{2}}$ & $9$  &  $\frac{2048 \pi^{14}}{40750666268358943771875}$   \\ \hline

  $\frac{17}{2};\frac{19}{2}$ & $30\sqrt{\frac{353452638}{a^{37}}}x^{\frac{17}{2}}(a-x)^{\frac{19}{2}}$ & $10$  &  $\frac{32768 \pi^{16}}{897125917897922147137828125}$   \\ \hline

 \hline\hline 
\end{tabular}

\end{threeparttable}
\label{tabelasume3}
\end{table}
\end{center}
\subsection{Using expression to obtain more infinite sums}
 Using the analytical expression from paper \cite{nis} 
\begin{equation}
\sum^{\infty}_{n=1}\frac{J_p(\frac{n\pi}{2})J_q(\frac{n\pi}{2})}{n^{p+q}}=\frac{\Gamma(p+q)(\frac{\pi}{4})^{p+q-\frac{1}{2}}}{\Gamma(p+q+\frac{1}{2})\Gamma(p+\frac{1}{2})\Gamma(q+\frac{1}{2})}-\frac{1}{2}\frac{(\frac{\pi}{4})^{p+q}}{\Gamma(p+1)\Gamma(q+1)} \label{nis1},
\end{equation}
it is possible to obtain even more infinite sums. For instance,
\begin{equation}
\sum^{\infty}_{n=1}\frac{J_1^2(\frac{n\pi}{2})}{n^2}=\frac{2}{3}-\frac{\pi^2}{32},
\end{equation}
whence, using previously calculated sum (\ref{prvsum}), we can obtain infinite sum 
\begin{equation}
\sum^{\infty}_{k=1}\frac{J_p^2(k\pi)}{(2k)^{2p}}=\frac{1}{3}-\frac{\pi^2}{32}.
\end{equation}
Further, using relation (\ref{nis1}) and results from Tables \ref{tabelasume} and \ref{tabelasume2}, we obtain new infinite sums quoted in Tables \ref{tabelasume4} and \ref{tabelasume5}, respectively.

\begin{widetext}
\begin{center}
\begin{table}
 
\caption{List of the infinite series $\sum^{\infty}_{k=1}\frac{J_p^2\left(k \pi \right)}{(2k)^{2p}}$.}
\begin{threeparttable}
 \begin{tabular}{|c |c| } 

 \hline\hline
  p & sum of the infinite series \\ [0.5ex] 
 \hline\hline \xrowht[()]{10pt}
 $1$ &  $\frac{1}{3}-\frac{\pi^2}{32}$  \\  \hline \xrowht[()]{10pt}
 
 $2$  &  $\frac{2\pi^2}{315}-\frac{\pi^4}{2048}$   \\ \hline \xrowht[()]{10pt}

 $3$  &  $\frac{8\pi^4}{155925}-\frac{\pi^6}{294912}$   \\ \hline \xrowht[()]{10pt}

 $4$  &  $\frac{16\pi^6}{70945875}-\frac{\pi^8}{75497472}$   \\ \hline \xrowht[()]{10pt}

$5$  &  $\frac{128\pi^8}{206239658625}-\frac{\pi^{10}}{30198988800}$   \\ \hline \xrowht[()]{10pt}

 $6$  &  $\frac{256\pi^{10}}{219150261254925}-\frac{\pi^{12}}{17394617548800}$   \\ \hline \xrowht[()]{10pt}

  $7$  &  $\frac{1024\pi^{12}}{641014514170655625}-\frac{\pi^{14}}{13637380158259200}$   \\ \hline \xrowht[()]{10pt}

 $8$  &  $\frac{2048\pi^{14}}{1234868674798755871875}-\frac{\pi^{16}}{13964677282057420800}$   \\ \hline \xrowht[()]{10pt}

   $9$  &  $\frac{32768\pi^{16}}{24246646429673571544265625}-\frac{\pi^{18}}{18098221757546417356800}$   \\ \hline \xrowht[()]{10pt}

   $10$  &  $\frac{65536\pi^{18}}{73863367240262256781014515625}-\frac{\pi^{20}}{28957154812074267770880000}$   \\ \hline

 \hline\hline 
\end{tabular}

\end{threeparttable}
\label{tabelasume4}
\end{table}
\end{center}
\end{widetext}

\begin{widetext}
\begin{center}
\begin{table}
 
\caption{List of the infinite series $\sum^{\infty}_{k=1}\frac{J_p\left(k \pi \right)J_q(k \pi)}{(2k)^{p+q}}$.}
\begin{threeparttable}
 \begin{tabular}{|c |c| } 

 \hline\hline
  p;q & sum of the infinite series \\ [0.5ex] 
 \hline\hline \xrowht[()]{10pt}
 $1;2$ &  $\frac{2\pi}{45}-\frac{\pi^3}{256}$  \\  \hline \xrowht[()]{10pt}
 
 $2;3$  &  $\frac{8\pi^3}{14175}-\frac{\pi^5}{24576}$   \\ \hline \xrowht[()]{10pt}

 $3;4$  &  $\frac{16\pi^5}{4729725}-\frac{\pi^7}{4718592}$   \\ \hline \xrowht[()]{10pt}

 $4;5$  &  $\frac{128\pi^7}{10854718875}-\frac{\pi^9}{1509949440}$   \\ \hline \xrowht[()]{10pt}

$5;6$  &  $\frac{256\pi^9}{9528272228475}-\frac{\pi^{11}}{724775731200}$   \\ \hline \xrowht[()]{10pt}

 $6;7$  &  $\frac{1024\pi^{11}}{23741278302616875}-\frac{\pi^{13}}{487049291366400}$   \\ \hline \xrowht[()]{10pt}

  $7;8$  &  $\frac{2048\pi^{13}}{39834473380605028125}-\frac{\pi^{15}}{436396165064294400}$   \\ \hline \xrowht[()]{10pt}

 $8;9$  &  $\frac{32768\pi^{15}}{692761326562102044121875}-\frac{\pi^{17}}{502728382154067148800}$   \\ \hline \xrowht[()]{10pt}

   $9;10$  &  $\frac{65536\pi^{17}}{1893932493340057866179859375}-\frac{\pi^{19}}{723928870301856694272000}$   \\ \hline

 \hline\hline 
\end{tabular}

\end{threeparttable}
\label{tabelasume5}
\end{table}
\end{center}
\end{widetext}

In a similar manner, starting from the relation 
\begin{equation}
\sum^{\infty}_{n=1}\frac{J_p(\frac{n \pi}{2})J_q(\frac{n \pi}{2})}{n^{p+q-2}}=\frac{(\frac{\pi}{4})^{p+q-\frac{5}{2}}\Gamma(p+q-2)}{2\Gamma(p-\frac{1}{2})\Gamma(q-\frac{1}{2})\Gamma(p+q-\frac{1}{2})}
\end{equation}
from the paper \cite{nis}, we obtain Table \ref{tabelasume6}.

\begin{widetext}
\begin{center}
\begin{table}
 
\caption{List of the infinite series $\sum^{\infty}_{k=0}\frac{J^2_p\left(\frac{(2k+1)\pi}{2} \right)}{(2k+1)^{2p-2}}$.}
\begin{threeparttable}
 \begin{tabular}{|c |c| } 

 \hline\hline
  p & sum of the infinite series \\ [0.5ex] 
 \hline\hline \xrowht[()]{10pt}
 $2$ &  $\frac{1}{15}$  \\  \hline \xrowht[()]{10pt}
 
 $3$  &  $\frac{2\pi^2}{2835}$   \\ \hline \xrowht[()]{10pt}

 $4$  &  $\frac{8\pi^4}{2027025}$  \\ \hline \xrowht[()]{10pt}

 $5$  &  $\frac{16\pi^6}{1206079875}$   \\ \hline \xrowht[()]{10pt}

$6$  &  $\frac{128\pi^8}{4331032831125}$   \\ \hline \xrowht[()]{10pt}

 $7$  &  $\frac{256\pi^{10}}{5478756531373125}$   \\ \hline \xrowht[()]{10pt}

  $8$  &  $\frac{1024\pi^{12}}{18589420910949013125}$   \\ \hline \xrowht[()]{10pt}

 $9$  &  $\frac{2048\pi^{14}}{40750666268358943771875}$   \\ \hline \xrowht[()]{10pt}

   $10$  &  $\frac{32768\pi^{16}}{897125917897922147137828125}$   \\ \hline

 \hline\hline 
\end{tabular}

\end{threeparttable}
\label{tabelasume6}
\end{table}
\end{center}
\end{widetext}

\subsection{Infinite series involving generalized hypergeometric functions}
The only case we have not discussed yet is the one where $\alpha$ is an integer and $\beta$ is a half-integer number. We start with $\alpha=1$, $\beta=\frac{1}{2}$, i.e. $\psi(x)=\frac{2\sqrt{3}}{a^2}x(a-x)^{\frac{1}{2}}$. Since $\alpha \neq \beta$, we again employ (\ref{kljucno2}) which leads to 
\begin{eqnarray}
C_n&=& \frac{2\sqrt{6}}{a^\frac{5}{2}}\frac{a^\frac{5}{2}}{2\i}B\left(\frac{3}{2},2\right)\left[_1\hspace{-0.05cm}F_1\left(2;\frac{7}{2};\i n\pi\right)-_1\hspace{-0.12cm}F_1\left(2;\frac{7}{2};-\i n\pi\right)\right] \nonumber \\
&=& \frac{\sqrt{6}}{\i}\frac{4}{15}\left[\sum^{\infty}_{k=0}\frac{(2)_k}{k!(\frac{7}{2})_k}(\i n\pi)^k-\sum^{\infty}_{k=0} \frac{(2)_k}{k!(\frac{7}{2})_k}(-\i n\pi)^k\right]\nonumber\\
&=& \frac{8\sqrt{6}}{15}n\pi\sum^{\infty}_{m=0}\frac{(2)_{2m+1}}{(2m+1)!(\frac{7}{2})_{2m+1}}(-n^2\pi^2)^m,
\end{eqnarray}
where $(a)_k=\frac{\Gamma(a+k)}{\Gamma(k)}$ are Pochhammer's symbols.
From that, using simple algebraic manipulations and the Legendre duplication formula $\Gamma(z)\Gamma(z+\frac{1}{2})=2^{1-2z}\sqrt{\pi}\Gamma(2z)$, we get
\begin{equation}
   C_n=\frac{8\sqrt{6}}{15}\frac{4}{7}n\pi\, _1\hspace{-0.05cm}F_2\left(2;\frac{9}{4},\frac{11}{4};-\frac{n^2\pi^2}{4}\right),
\end{equation}
whence from condition $\sum^{\infty}_{n=1}|C_n|^2=1$ we obtain
\begin{equation}
 \sum^{\infty}_{n=1}n^2
 \left[{_1\hspace{-0.05cm}F_2}\left(2;\frac{9}{4},\frac{11}{4};-\frac{n^2\pi^2}{4}\right)\right]^2=\frac{3675}{2048\pi^2}.
\end{equation}
With the same procedure, we get the Table \ref{tabelasume7}. Similar to the case $\alpha=\beta$, for each wave function we obtain one infinite sum. Therefore, the order of computing infinite series from Table \ref{tabelasume7} is irrelevant.

\begin{widetext}
\begin{center}
\begin{table}
 
\caption{List of the infinite series that can be calculated for integer $\alpha$ and half-integer $\beta$.}
\begin{threeparttable}
 \begin{tabular}{|c |c| c |c|} 

 \hline\hline
 $\alpha$,$\beta$ & $\psi(x)$  & infinite series & sum of the infinite series \\ [0.5ex] 
 \hline\hline \xrowht[()]{14pt}
 $1$, $\frac{1}{2}$ & $\frac{2\sqrt{3}}{a^2}x(a-x)^{\frac{1}{2}}$  & $\sum^{\infty}_{n=1}n^2
 \left[{_1\hspace{-0.05cm}F_2}(2;\frac{9}{4},\frac{11}{4};-\frac{n^2\pi^2}{4})\right]^2$ &  $\frac{3675}{2048\pi^2}$ \\ \hline \xrowht[()]{14pt}
 
 $2$, $\frac{1}{2}$ & $\frac{\sqrt{30}}{a^3}x^2(a-x)^{\frac{1}{2}}$ & $\sum^{\infty}_{n=1}n^2
 \left[{_2\hspace{-0.05cm}F_3}(2,\frac{5}{2};\frac{3}{2},\frac{11}{4},\frac{13}{4};-\frac{n^2\pi^2}{4})\right]^2$  &  $\frac{6615}{4096\pi^2}$   \\ \hline \xrowht[()]{14pt}

 $3$, $\frac{1}{2}$ & $2\frac{\sqrt{14}}{a^4}x^3(a-x)^{\frac{1}{2}}$ & $\sum^{\infty}_{n=1}n^2
 \left[{_2\hspace{-0.05cm}F_3}(3,\frac{5}{2};\frac{3}{2},\frac{13}{4},\frac{15}{4};-\frac{n^2\pi^2}{4})\right]^2$  &  $\frac{1715175}{1048576\pi^2}$   \\ \hline \xrowht[()]{14pt}

 $1$,$\frac{3}{2}$ & $2\frac{\sqrt{15}}{a^3}x(a-x)^{\frac{3}{2}}$ & $\sum^{\infty}_{n=1}n^2
 \left[{_1\hspace{-0.05cm}F_2}(2;\frac{11}{4},\frac{13}{4};-\frac{n^2\pi^2}{4})\right]^2$  &  $\frac{6615}{2048\pi^2}$   \\ \hline \xrowht[()]{14pt}

$2$, $\frac{3}{2}$ & $2\frac{\sqrt{70}}{a^4}x^2(a-x)^{\frac{3}{2}}$ & $\sum^{\infty}_{n=1}n^2
 \left[{_2\hspace{-0.05cm}F_3}(2,\frac{5}{2};\frac{3}{2},\frac{13}{4},\frac{15}{4};-\frac{n^2\pi^2}{4})\right]^2$  &  $\frac{38115}{16384\pi^2}$   \\ \hline \xrowht[()]{14pt}

 $3$,$\frac{3}{2}$ & $2\frac{\sqrt{210}}{a^5}x^3(a-x)^{\frac{3}{2}}$ & $\sum^{\infty}_{n=1}n^2
 \left[{_2\hspace{-0.05cm}F_3}^2(3,\frac{5}{2};\frac{3}{2},\frac{15}{4},\frac{17}{4};-\frac{n^2\pi^2}{4})\right]^2$  &  $\frac{2147145}{1048576\pi^2}$   \\ \hline \xrowht[()]{14pt}

 $1$, $\frac{5}{2}$ & $2\frac{\sqrt{42}}{a^4}x(a-x)^{\frac{5}{2}}$ & $\sum^{\infty}_{n=1}n^2
 \left[{_1\hspace{-0.05cm}F_2}(2;\frac{13}{4},\frac{15}{4};-\frac{n^2\pi^2}{4})\right]^2$  &  $\frac{22869}{4096\pi^2}$   \\ \hline \xrowht[()]{14pt}

 $2$, $\frac{5}{2}$ & $6\frac{\sqrt{35}}{a^5}x^2(a-x)^{\frac{5}{2}}$ & $\sum^{\infty}_{n=1}n^2
 \left[{_2\hspace{-0.05cm}F_3}(2,\frac{5}{2};\frac{3}{2},\frac{15}{4},\frac{17}{4};-\frac{n^2\pi^2}{4})\right]^2$  &  $\frac{143143}{40960\pi^2}$   \\ \hline \xrowht[()]{14pt}

  $3$, $\frac{5}{2}$ & $6\frac{\sqrt{154}}{a^3}x^3(a-x)^{\frac{5}{2}}$ &  $\sum^{\infty}_{n=1}n^2
 \left[{_2\hspace{-0.05cm}F_3}(3,\frac{5}{2};\frac{3}{2},\frac{17}{4},\frac{19}{4};-\frac{n^2\pi^2}{4})\right]^2$ &  $\frac{2927925}{1048576 \pi^2}$   \\ \hline

 \hline\hline 
\end{tabular}

\end{threeparttable}
\label{tabelasume7}
\end{table}
\end{center}
\end{widetext}

\begin{figure}[ht] 
\includegraphics[width=12.4cm]{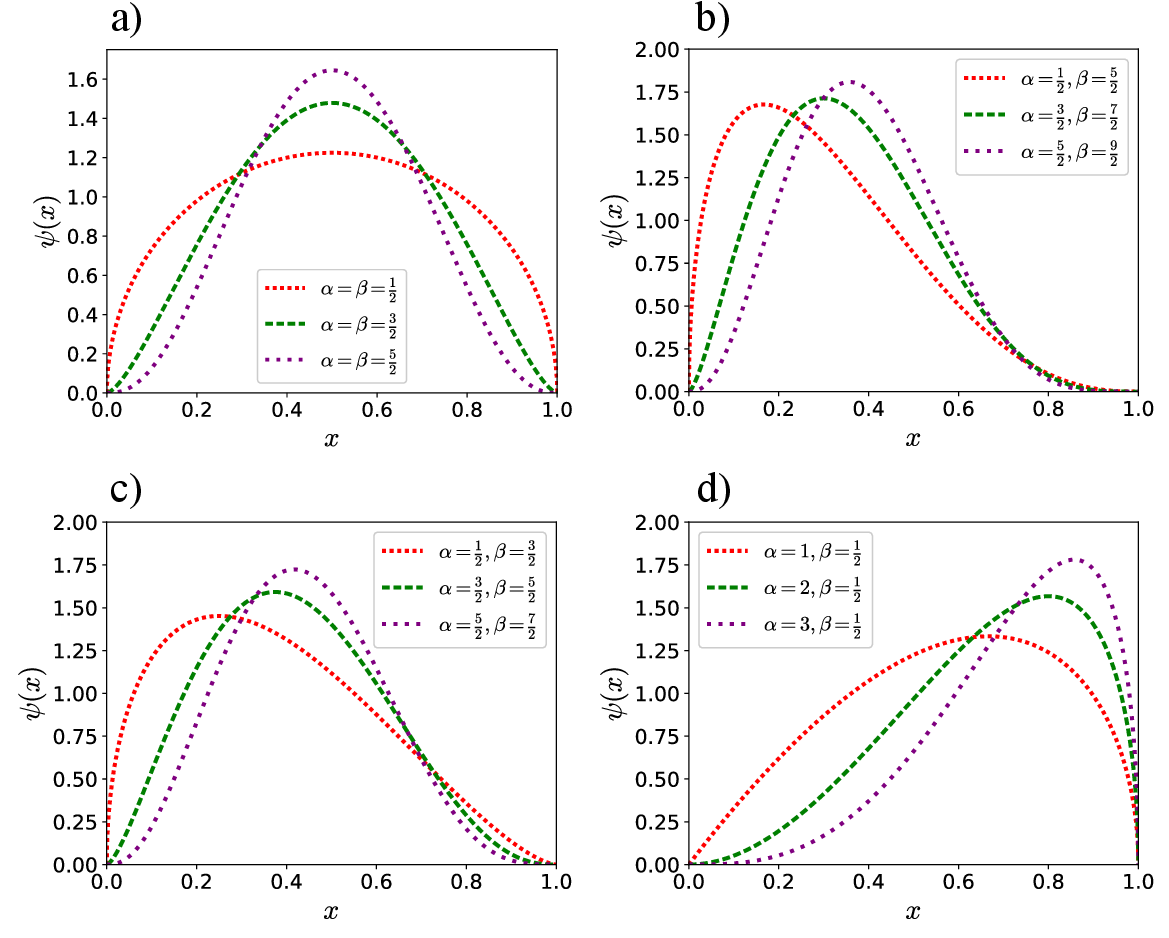}
\centering
\caption{\label{Fig1}
(Color online) The first three wave functions $\psi(x)$ for $a=1$ from Table \ref{tabelasume} (a)), Table \ref{tabelasume2} (b)), Table \ref{tabelasume3} (c)), and Table \ref{tabelasume7} (d)).}
\end{figure}
In Fig \ref{Fig1} the first three wave functions $\psi(x)$ from Tables \ref{tabelasume},\ref{tabelasume2},\ref{tabelasume3} and \ref{tabelasume7} are presented. All wave functions investigated in the paper contain zero nodes, which implies that measurement of higher values of energies in these types of states is less probable.  When $\alpha=\beta=\frac{i}{2}$, $i$ is an integer (see Fig \ref{Fig1} a)), the wave functions are symmetric around line $x=a/2$, which explains the absence of even terms in the expansion (\ref{linearnakom}) in this case, tremendously simplifying the calculations of sums. In Figs \ref{Fig1} b), c), and d) this symmetry is not present and consequently both even and odd terms in (\ref{linearnakom}) appear. In Figs \ref{Fig1} b) and c) one can see that with increasing values of $\alpha$ and $\beta$ we come closer to the aforementioned symmetry, which is in contrast to Fig \ref{Fig1} d). 

\subsection{General summation formulas}
Instead of using \eqref{kljucno} for $\alpha=\beta$ and \eqref{kljucno2} for $\alpha\neq\beta$, one can generate infinite sums $\sum^{\infty}_{k=0}\frac{J_p^2 (\frac{(2k+1) \pi}{2})}{(2k+1)^{2p}}$, $\sum^{\infty}_{k=1}\frac{J_p^2 (k \pi)}{(2k)^{2p-2}}$, $\sum^{\infty}_{k=0}\frac{J_p (\frac{(2k+1) \pi}{2})J_q (\frac{(2k+1) \pi}{2})}{(2k+1)^{p+q}}$, and $\sum^{\infty}_{k=1}k^2
\left[{_p\hspace{-0.05cm}F_{p+1}}(\alpha_1,\alpha_2,...,\alpha_p;\beta_1,\beta_2,...,\beta_{p+1};-\frac{k^2\pi^2}{4})\right]^2$, where $p$, $q$ are integers, by employing one general relation, which will be derived in this subsection. We will also obtain general formulas of the similar type by twofold calculations of average values of $\hat{H}$ and $\hat{H}^2$. 
 
 Without loss of generality, we set $\frac{\hbar^2}{2m}=a=1$ throughout this subsection.
Consider the states 
\begin{equation}
\psi_{\alpha \beta}(x) = C_{\alpha \beta} x^\alpha (1-x)^\beta,
\end{equation}
where $\alpha, \beta\geq 1/2$. The normalization constant $C_{\alpha \beta}$ is determined from
\begin{equation}
\int_0^1 \;[\psi_{\alpha \beta}(x)]^2\mathrm{d}x = 1.
\end{equation}
To obtain the value for $C_{\alpha \beta}$,
one needs to calculate the integral
\begin{equation}
K_{\alpha \beta} = \int_0^1  x^{2 \alpha} (1-x)^{2 \beta}\mathrm{d}x.
\end{equation}
Since
\begin{equation}
(1-x)^{2 \beta} = \sum_{k=0}^\infty \frac{(-2 \beta)_k}{k!}x^{k},
\end{equation}
we find
\begin{equation}
K_{\alpha \beta} = \sum_{k=0}^\infty \frac{(-2 \beta)_k}{k!} \frac{1}{2 \alpha +k+1}
= \sum_{k=0}^\infty \frac{(-2 \beta)_k}{k!} \frac{\Gamma(2 \alpha +1 + k)}{\Gamma(2\alpha +2+k )}.
\end{equation}
Thus,
\begin{equation}
    K_{\alpha \beta} = \frac{\Gamma(2 \alpha +1)}{\Gamma(2 \alpha +2)} \sum_{k=0}^\infty
    \frac{(-2 \beta)_k}{k!} \frac{(2 \alpha +1)_k}{(2 \alpha +2)_k}
    = \frac{\Gamma(2 \alpha +1)}{\Gamma(2 \alpha +2)}\, _2\hspace{-0.05cm} F_1\left( -2 \beta, 2 \alpha +1; 2 \alpha +2;1  \right)
\end{equation}
and, by the Gauss theorem,
\begin{equation}
K_{\alpha \beta} = \frac{\Gamma(2 \alpha +1) \Gamma(2 \beta +1)}{\Gamma(2 \alpha + 2 \beta +2)}=B(2\alpha+1,2\beta+1).
\label{Kab}
\end{equation}
Hence, normalized states are given by
\begin{equation}
  \psi_{\alpha \beta}(x) = \sqrt{\frac{1}{B(2\alpha+1,2\beta+1)}}\,x^\alpha (1-x)^\beta \label{Phiab}.
\end{equation}
The states $\psi_{\alpha \beta}(x)$ may be expanded in terms of the basis states
$\psi_n(x) = \sqrt{2} \sin(n \pi x)$ as
\begin{equation}
     \psi_{\alpha \beta}(x) = \sum_{n=1}^\infty C_{\alpha \beta}^{(n)}\psi_n(x),
\end{equation}
where
\begin{equation}
C_{\alpha \beta}^{(n)} = \int_0^1  \;\psi_{\alpha \beta}(x) \psi_n(x)\mathrm{d}x
= \sqrt{\frac{2}{K_{\alpha \beta}}} I_{\alpha \beta}(n \pi)
\end{equation}
and
\begin{equation}
I_{\alpha \beta} (b) =  \int_0^1  \; x^\alpha (1-x)^\beta \sin(b x)\mathrm{d}x.
\end{equation}
The integral $I_{\alpha \beta} (b)$ may be evaluated in a similar manner as the 
integral $K_{\alpha \beta}$, leading to
\begin{equation}
\begin{split}
 I_{\alpha \beta}(b) & = bB(\alpha+2,\beta+1) \times \\
 & \times  _2 \hspace{-0.13cm} F _3  \left( \frac{\alpha +2}{2}, \frac{\alpha +3}{2};
 \frac{\alpha+ \beta +3}{2}, \frac{\alpha+ \beta +4}{2}, \frac 32; - \frac{b^2}{4}\right).
\end{split}
\end{equation}
The condition $\sum_{n=1}^\infty |C_{\alpha \beta}^{(n)}|^2 = 1$ now gives the summation formula for hypergeometric functions
\begin{equation}
\begin{split}
& \sum_{n=1}^\infty n^2 \left[ _2 \hspace{-0.05cm} F _3  \left( \frac{\alpha +2}{2}, \frac{\alpha +3}{2};
 \frac{\alpha +\beta +3}{2}, \frac{\alpha +\beta +4}{2}, \frac 32; - \frac{n^2\pi^2}{4}\right)  \right]^2 \\
 & = \frac{1}{2\pi^2}\left[\frac{1}{B(\alpha+2,\beta+1)}\right]^2B(2\alpha+1,2\beta+1)\\
 &= \frac{1}{2\pi^2}\left[\frac{\Gamma(\alpha+\beta+3)}{\Gamma(\alpha+2)\Gamma(\beta+1)}\right]^2\frac{\Gamma(2\alpha+1)\Gamma(2\beta+1)}{\Gamma(2\alpha+2\beta+2)}, \label{generalka}
\end{split}    
\end{equation}
which is valid for all $\alpha,\beta \geq 1/2$. This represents a general relation from which sums of infinite series obtained in tables \ref{tabelasume}, \ref{tabelasume2}, \ref{tabelasume3}, and \ref{tabelasume7} can also be generated. For instance,
\begin{equation}
    \sum^{\infty}_{k=0}\frac{J_p^2\left(\frac{(2k+1)\pi}{2}\right)}{(2k+1)^{2p}}=\frac{\pi^{2p-2}}{2^{3-4p}}\frac{\left[\Gamma(p)\right]^2}{\Gamma(4p)}
\end{equation}
is obtained from \eqref{generalka} for $\alpha=\beta=\frac{2r+1}{2}$, $p=r+1$, and $r=0,1,2,...$ (see Table \ref{tabelasume}).

Further, the mean (expectation) value of energy in the given state $\psi_{\alpha \beta}(x)$ can be calculated by using both
\begin{equation}
\langle \hat{H} \rangle_{\psi_{\alpha \beta}}=\int^1_0 \psi_{\alpha \beta}^*(x)\left(-\frac{\d^2}{\d x^2}\right)\psi_{\alpha \beta}(x)\d x \label{srednjajedan}, 
\end{equation}
and
\begin{equation}
\langle \hat{H} \rangle_{\psi_{\alpha \beta}}=\sum^{\infty}_{n=1}|C^{(n)}_{\alpha \beta}|^2E_n=\sum^{\infty}_{n=1}|C^{(n)}_{\alpha \beta}|^2 n^2\pi^2,\label{srednjadva}
\end{equation}
where 
\begin{equation}
|C^{(n)}_{\alpha \beta}|^2=\Bigg|\int^1_0 \psi_n(x) \psi_{\alpha \beta}(x)\d x\Bigg|^2.
\label{verovatnoca}\end{equation}
Following the same procedure, from \eqref{srednjajedan} we obtain
\begin{equation}
  \langle \hat{H} \rangle_{\psi_{\alpha \beta}}=\frac{\alpha \beta B(2\alpha-1,2\beta-1)}{(2\alpha+2\beta-1)B(2\alpha+1,2\beta+1)}, \quad \alpha, \beta>\frac{1}{2}, \label{wolslo} 
\end{equation}
whereas from \eqref{srednjadva} we get
\begin{equation}
  \langle \hat{H} \rangle_{\psi_{\alpha \beta}}=\sum^{\infty}_{n=1}\frac{2\pi^4B^2(\alpha+2,\beta+1)}{B(2\alpha+1,2\beta+1)}n^4 \left[\,_2 \hspace{-0.05cm} F _3  \left( \frac{\alpha +2}{2}, \frac{\alpha +3}{2};
 \frac{\alpha+ \beta +3}{2}, \frac{\alpha+ \beta +4}{2}, \frac 32; - \frac{n^2\pi^2}{4}\right)\right]^2. \label{wolslo2} 
\end{equation}
Since according to quantum mechanics the two ways of energy averaging are equivalent,
equations \eqref{wolslo} and \eqref{wolslo2} give identical results, and from that
\begin{align}
   &\sum^{\infty}_{n=1}n^4 \left[\,_2 \hspace{-0.05cm} F _3  \left( \frac{\alpha +2}{2}, \frac{\alpha +3}{2};
 \frac{\alpha+ \beta +3}{2}, \frac{\alpha+ \beta +4}{2}, \frac 32; - \frac{n^2\pi^2}{4}\right)\right]^2=\frac{\alpha \beta B(2\alpha-1,2\beta-1)}{2\pi^4 (2\alpha+2\beta-1) B^2(\alpha+2,\beta+1)}\nonumber \\&=\frac{\alpha\beta\Gamma(2\alpha-1)\Gamma(2\beta-1)}{2\pi^{4}(2\alpha+2\beta-1)\Gamma(2\alpha+2\beta-2)}\left[\frac{\Gamma(\alpha+\beta+3)}{\Gamma(\alpha+2)\Gamma(\beta+1)}\right]^{2}, \quad \alpha,\beta>\frac{1}{2}.
\end{align}
From here directly, without using \cite{nis}, we can obtain Table \ref{tabelasume6} for half-integer values of $\alpha=\beta \geq \frac{3}{2}$, and for $\alpha \neq \beta$ potentially new sums of infinite series. Moreover, sums of all infinite series from \cite{mali} can be evaluated from this relation for integer values of $\alpha$ and $\beta$ as well.

Finally, following paper \cite{ajp}, we start from
\begin{equation}
    \langle \hat{H}^{2} \rangle_{\psi_{\alpha \beta}}=\left(\hat{H}\psi_{\alpha\beta},\hat{H}\psi_{\alpha\beta}\right)=\sum^{\infty}_{n=1}|C^{(n)}_{\alpha \beta}|^2E_n^{2}=\sum^{\infty}_{n=1}|C^{(n)}_{\alpha \beta}|^2 n^4\pi^4. \label{glup1}
\end{equation}
On the other hand
\begin{align}
    \langle \hat{H}^{2} \rangle_{\psi_{\alpha \beta}}&=\int^1_0 \frac{\d^2}{\d x^2}\psi_{\alpha \beta}^*(x)\frac{\d^2}{\d x^2}\psi_{\alpha \beta}(x)\d x\nonumber \\ &=\frac{3\alpha\beta(\beta-1)\Gamma(2\alpha-1)\Gamma(2\beta-3)}{2(2\alpha-3)(2\alpha+2\beta-5)(2\alpha+2\beta-3)\Gamma(2\alpha+2\beta-6)B(2\alpha+1,2\beta+1)}, \quad \alpha,\beta>\frac{3}{2}. \label{glup2}
\end{align}
From \eqref{glup1} and \eqref{glup2} we obtain 
\begin{align}
 &\sum^{\infty}_{n=1}n^6 \left[\,_2 \hspace{-0.05cm} F _3  \left( \frac{\alpha +2}{2}, \frac{\alpha +3}{2};
 \frac{\alpha+ \beta +3}{2}, \frac{\alpha+ \beta +4}{2}, \frac 32; - \frac{n^2\pi^2}{4}\right)\right]^2 \nonumber \\ &=\frac{3\alpha\beta(\beta-1)\Gamma(2\alpha-1)\Gamma(2\beta-3)}{4\pi^{6}(2\alpha-3)(2\alpha+2\beta-5)(2\alpha+2\beta-3)\Gamma(2\alpha+2\beta-6)}\left[\frac{\Gamma(\alpha+\beta+3)}{\Gamma(\alpha+2)\Gamma(\beta+1)}\right]^{2}, \quad \alpha,\beta>\frac{3}{2},    
\end{align}
which allows us to compute sums of new infinite series. For instance, by choosing $\alpha=\beta=\frac{5}{2}$ we obtain
\begin{equation}
    \sum^{\infty}_{k=0}\frac{J_{3}^{2}\left(\frac{(2k+1)\pi}{2}\right)}{(2k+1)^{2}}=\frac{1}{35}.
\end{equation}

\section{Conclusion}

In this paper, we have shown how a simple physical model of a particle trapped inside an infinite quantum potential well can be exploited in order to calculate infinite sums of various nontrivial infinite series analytically. For that purpose, we use different non-polynomial and non-trigonometric wave functions, which allow us to evaluate the infinite sums of which some are presented in Tables \ref{tabelasume}, \ref{tabelasume2}, \ref{tabelasume3} and \ref{tabelasume7}. Furthermore, by using results from \cite{nis}, we also obtain related sums, examples of which are given in Tables \ref{tabelasume4}, \ref{tabelasume5}, and \ref{tabelasume6}. Note that this method allows us to calculate infinitely many sums of infinite series in closed form. These are the series of the type: $\sum^{\infty}_{k=0}\frac{J_p^2 (\frac{(2k+1) \pi}{2})}{(2k+1)^{2p}}$,$\sum^{\infty}_{k=1}\frac{J_p^2 (k \pi)}{(2k)^{2p-2}}$, $\sum^{\infty}_{k=0}\frac{J_p (\frac{(2k+1) \pi}{2})J_q (\frac{(2k+1) \pi}{2})}{(2k+1)^{p+q}}$, $\sum^{\infty}_{k=1}\frac{J_p^2 (k \pi)}{(2k)^{2p}}$, $\sum^{\infty}_{k=1}\frac{J_p (k \pi)J_q (k \pi)}{(2k)^{p+q}}$, $\sum^{\infty}_{k=0}\frac{J_p^2 (\frac{(2k+1) \pi}{2})}{(2k+1)^{2p-2}}$, $\sum^{\infty}_{k=1}k^2
\left[{_p\hspace{-0.05cm}F_{p+1}}(\alpha_1,\alpha_2,...,\alpha_p;\beta_1,\beta_2,...,\beta_{p+1};-\frac{k^2\pi^2}{4})\right]^2$, where $p$, $q$ are integers and some values of $\alpha_i$, $\beta_i$ are given in Table \ref{tabelasume7}. Moreover, we derive three general relations from which all aforementioned sums of infinite series as well as the new ones can be computed. As far as we are aware, these sums have not been obtained in closed form previously. Due to the fact that the starting quantum mechanical problem is well-defined, the series convergence is secured by the properties of the corresponding Hilbert space.  Since the method is geometric in origin and is based on the properties of Hilbert space, by using the analogous type of procedure in some other well-defined quantum mechanical problems, various new infinite sums could potentially be evaluated. 

\section{Appendix} \label{app}

\subsection{First integral}

\renewcommand{\theequation}{A.\arabic{equation}}
\setcounter{equation}{0}

Starting from the integral representation of the Bessel function of the first kind
\begin{equation}
 J_{\nu}(z)=\frac{1}{\Gamma(\nu+\frac{1}{2})\sqrt{\pi}}\left(\frac{z}{2}\right)^{\nu}\int^1_{-1}(1-t^2)^{\nu-\frac{1}{2}}\e^{\mathrm{i} zt}\d t, \quad \mathrm{Re}\,\nu>-\frac{1}{2}
\end{equation}
by putting $\nu=\mu-\frac{1}{2}$, $z=\frac{\tilde{a}u}{2}$, we get
\begin{equation}
 J_{\mu-\frac{1}{2}}\left(\frac{\tilde{a}u}{2}\right)=\frac{1}{\Gamma(\mu)\sqrt{\pi}}\left(\frac{\tilde{a}u}{4}\right)^{\mu-\frac{1}{2}}\int^1_{-1}(1-t^2)^{\mu-1}\e^{\mathrm{i} \frac{\tilde{a}u}{2}t}\d t, \quad \mathrm{Re}\,\mu>0
\end{equation}
which can also be written as
\begin{equation}
 \Gamma(\mu)\sqrt{\pi}J_{\mu-\frac{1}{2}}\left(\frac{\tilde{a}u}{2}\right)=\left(\frac{\tilde{a}u}{4}\right)^{\mu-\frac{1}{2}}\int^1_{-1}(1-t^2)^{\mu-1}\cos\left(\frac{\tilde{a}u}{2}t\right)\d t, \quad \mathrm{Re}\,\mu>0.  
\end{equation}
Using substitution $ut=l$, we obtain
\begin{equation}
 \Gamma(\mu)\sqrt{\pi}J_{\mu-\frac{1}{2}}\left(\frac{\tilde{a}u}{2}\right)=\left(\frac{\tilde{a}}{4u}\right)^{\mu-\frac{1}{2}}\int^u_{-u}\cos\left(\frac{\tilde{a}l}{2}\right)(u-l)^{\mu-1}(u+l)^{\mu-1}\d l.
\end{equation}
Substituting $u+l=2x$, we get
\begin{equation}
 \Gamma(\mu)\sqrt{\pi}J_{\mu-\frac{1}{2}}\left(\frac{\tilde{a}u}{2}\right)=\left(\frac{\tilde{a}}{u}\right)^{\mu-\frac{1}{2}}\int^u_{0}\cos\left(\frac{\tilde{a}}{2}(2x-u)\right)(u-x)^{\mu-1}x^{\mu-1}\d x.   
\end{equation}
Multiplying both sides by $\sin(\frac{\tilde{a}u}{2})$ and employing elementary transformations leads to \eqref{kljucno}.
\subsection{Second integral}

\renewcommand{\theequation}{B.\arabic{equation}}
\setcounter{equation}{0}

Starting from
\begin{align}
&\int^u_0 x^{\nu-1}(u-x)^{\mu-1}\sin(\tilde{a}x)\d x=\int^u_0 x^{\nu-1}(u-x)^{\mu-1}\frac{\mathrm{e}^{\mathrm{i}\tilde{a}x}-\mathrm{e}^{-\mathrm{i}\tilde{a}x}}{2\mathrm{i}}\mathrm{d}x= \nonumber \\ &=
\frac{1}{2\mathrm{i}}\left(\int^u_0 x^{\nu-1}(u-x)^{\mu-1}\mathrm{e}^{\mathrm{i}\tilde{a}x}\mathrm{d}x\large-\int^u_0 x^{\nu-1}(u-x)^{\mu-1}\mathrm{e}^{-\mathrm{i}\tilde{a}x}\mathrm{d}x\right),
\end{align}
by employing substitution 
$x=ut$ and using the definition of beta function and integral representation of  confluent hypergeometric (Kummer's) function
\begin{equation}
_1\hspace{-0.5mm}F_1(\alpha;\beta;z)=\frac{\Gamma(\beta)}{\Gamma(\alpha)\Gamma(\beta-\alpha)}\int^1_0\mathrm{e}^{zt}t^{\alpha-1}(1-t)^{\beta-\alpha-1}\mathrm{d}t
\end{equation}
we obtain the identity \eqref{kljucno2}.

\begin{acknowledgments}

The authors gratefully acknowledge the financial support of the Ministry of Science, Technological Development and Innovation of the Republic of Serbia (Grants No. 
451-03-66/2024-03/200125 and 451-03-65/2024-03/200125).

\end{acknowledgments}


\end{document}